\documentclass[aps,pra,preprint,groupedaddress,showpacs]{revtex4-1}
\usepackage{feynmp}
\usepackage[pdftex]{graphicx}
\usepackage{braket}
\usepackage{amsbsy}
\usepackage{amsmath}
\usepackage{amssymb}
\usepackage{bm}
\usepackage{euscript}
\let\Im\undefined
\DeclareMathOperator{\Im}{Im}
\let\Re\undefined
\DeclareMathOperator{\Re}{Re}
\DeclareMathOperator{\Tr}{Tr}
\DeclareMathOperator{\arcctg}{arcctg}
\begin{document}

\title{Calculations of magnetic states and minimum energy paths of transitions using a noncollinear extension of the 
Alexander-Anderson model and a magnetic force theorem}

\author{Pavel F. Bessarab$^{1,2}$, Valery M. Uzdin$^{2,3}$, and  Hannes J\'onsson$^{1,4}$}
\affiliation{$^1$Science Institute and Faculty of Physical Sciences, VR-III, University of 
Iceland, 107 Reykjav\'{\i}k, Iceland}
\affiliation{$^2$Department of Physics, St. Petersburg State University, St. Petersburg, 198504, Russia}
\affiliation{$^3$St.Petersburg National Research University of Information Technologies, Mechanics and Optics, St. Petersburg, 197101, Russia }
\affiliation{$^4$Dept. of Applied Physics, Aalto University, Espoo,  FI-00076, Finland}

\date{\today}
\begin{abstract}
Calculations of stable and metastable magnetic states as well as minimum energy paths for transitions between states 
are carried out using a noncollinear extension of the
multiple-impurity Alexander-Anderson model and a 
magnetic force theorem which is derived and used to evaluate the total energy gradient with respect to 
orientation of magnetic moments -- an important tool for efficient navigation on the energy surface.
By using this force theorem, the search for stable and metastable magnetic states as well as minimum energy paths revealing the mechanism and activation energy of transitions can be carried out efficiently.
For Fe monolayer on W(110) surface, the model gives magnetic moment as well as exchange coupling between 
nearest and next-nearest neighbors that are in good agreement with previous density functional theory calculations. 
When applied to nanoscale Fe islands on this surface, the magnetic moment is predicted to be 10\% larger for atoms
at the island rim, explaining in part an experimentally observed trend in the energy barrier for magnetization reversal 
in small islands. 
Surprisingly, the magnetic moment of the atoms does not change much along the minimum energy path for the 
transitions, which for islands containing more than 15 atom rows along either $[001]$ or $[1\bar{1}0]$ 
directions involves the formation of a thin, temporary domain wall. A noncollinear magnetic state is identified in a 
$7\times 7$ atomic row Fe island where the magnetic moments are arranged in an antivortex configuration with the 
central ones pointing out of the $(110)$ plane. This illustrates how the model can describe complicated exchange 
interactions even though it contains only a few parameters. The minimum energy path between this antivortex state and the collinear ground state is also calculated and the thermal stability of the antivortex state estimated.
\end{abstract}


\maketitle


\section{Introduction}

In theoretical studies of stable and metastable states of magnetic systems, 
the determination of magnetic forces is often required, i.e. the total energy gradient with respect to the orientation of the 
magnetic moments. Such forces are needed to calculate the dynamics of the system and to guide a minimization of the 
energy to identify stable or metastable magnetic states. Moreover, magnetic forces are particularly important for studying 
thermally activated magnetic transitions~\cite{bessarab_12,fiedler_12}, where a minimum energy path (MEP) connecting 
the initial and final states
needs to be found. An MEP reveals the 
optimal mechanism of a transition, showing how each magnetic vector rotates so as to minimize the energy barrier
to the transition. 
The highest energy point
along the MEP represents a first order saddle point on the energy surface and can 
be used to estimate the activation energy for the transition. An efficient method for evaluating magnetic forces is 
essential for these kinds of calculations.  

Density functional theory (DFT) can be used to study a wide range of magnetic systems. In the spin polarized extension using a $2\times2$ spin-density matrix, noncollinear states can be calculated and characterized~\cite{vonBarth_72}. 
Several studies of stationary magnetic states in various types of systems, including noncollinear systems, 
have been carried out using this approach~\cite{bihlmayer_07}.
Within an adiabatic approximation, the magnitude of the magnetic moments as well as the electronic charge 
in atomic volumes are assumed to be determined by the instantaneous orientation of the magnetic moments. The magnitude and electronic charge are determined using self-consistent, quantum mechanical calculations, while the orientation is treated classically~\cite{antropov_96}. 
The magnetic force giving the change in the energy with 
the direction of the magnetic moments can be
calculated using a magnetic force theorem derived within the local spin density approximation (LSDA) to the exchange correlation functional~\cite{liechtenstein_84,oswald_85,liechtenstein_87}. This approach can, for example, be used to carry out simulations of spin dynamics using the 
Landau-Lifshitz equation of motion~\cite{antropov_96}. Such calculations are, however, quite involved and are 
computationally intensive. 
In order to describe nonstationary magnetic states, local constraints need to be introduced mimicking the effect of a local field acting on each magnetic moment so as to keep its orientation fixed 
in
the predefined direction~\cite{stocks_98}. This local field needs to be determined self-consistently. An initial guess for the constraint is made, the system then relaxed subject to this constraint, the calculated
orientation evaluated and compared with the predefined orientation, the constraint revised, etc. 
This adds an inner self-consistency loop to the DFT calculations. As a result, the calculation of arbitrary nonstationary, noncollinear arrangements of the magnetic moments using DFT is challenging.

Most calculations of spin dynamics and magnetic transitions are presently carried out using simple, phenomenological models, in particular Heisenberg-type models, where the magnitude of the magnetic moments is assumed to be constant upon rotation. The determination of the total energy and its gradient with respect to orientation of the magnetic moments is then straightforward. However, in order to describe magnetic systems accurately enough, the model Hamiltonian may need to include several phenomenological terms. In addition
to the usual magnetic exchange, anisotropy, dipole-dipole interaction, more elaborate interactions such as biquadratic exchange and Dzyaloshinsky-Moriya interaction have been invoked to better reproduce observed
properties of magnetic systems~\cite{heinze_11}.  
The magnitude of the magnetic moments and interaction parameters in model Hamiltonians can, in principle, be calculated using DFT. Typically, this is done for collinear  
states~\cite{liechtenstein_87,udvardi_03,daalderop_90,wu_12} and the parameters then kept the same for arbitrary 
noncollinear ordering of the magnetic moments. This approach can be accurate enough for small deviations from the collinear states, but is expected to fail for large rotation angles in itinerant electron systems~\cite{lounis_10} where the magnitude of magnetic moments and coupling parameters depend on the relative orientation of the moments. Sophisticated schemes have been proposed to make Heisenberg type models reproduce such behavior. 
Drautz and F\"ahnle~\cite{drautz_04} used a spin-cluster expansion to parametrize the total energy of a magnetic system as a function of orientation of the magnetic moments. 
Recently, Szilva {\it et al.}~\cite{szilva_13} derived an expression for the intersite exchange for an arbitrary alignment of spins making it easier to interpolate calculations for several points in configuration space.
While additional parameters and elaborate expressions for the dependence of the parameters on the orientation of the magnetic moments can in principle be used to make a Heisenberg type model fit a given system,  
the transferability of the parameter values obtained in this way 
can be of concern and the question arises whether a different model that only requires a few, well defined parameters
could 
be used instead. 

The Alexander-Anderson (AA) model~\cite{anderson_61,alexander_64}
generalized to include multiple magnetic impurities 
describes magnetic systems 
containing itinerant electrons.
It includes two electronic bands: a quasilocalized band of {\it d}-electrons and a band of itinerant {\it s}({\it p}) electrons. 
The model has been shown to give results that are consistent with DFT calculations but it also provides a clear physical picture of the basic properties of the systems studied~\cite{uzdin_12,uzdin_09a}.    
A noncollinear extension of the AA model (NCAA) has been developed
in mean-field approximation as well as an efficient implementation 
of the self-consistency calculations using the recursive Green function method~\cite{haydock_72,haydock_75} and analytical transformations of the density of states~\cite{uzdin_98,uzdin_09}.
This makes it possible to apply the NCAA model to 
large and complex magnetic systems where a self-consistent calculation of the number of {\it d} electrons and magnitude of the magnetic moments is carried out for a fixed orientation of the magnetic moments.
NCAA has, for example, been used successfully to describe magnetism of 3{\it d}-metal surfaces and interfaces~\cite{uzdin_09}. Moreover,  
a noncollinear ordering of magnetic moments in nanoclusters of 3{\it d}-metal atoms was 
obtained in
calculations using the NCAA model~\cite{uzdin_99,uzdin_01,uzdin_00}, and this 
prediction
was later confirmed by DFT calculations~\cite{gotsis_06,bergman_06}.

For an arbitrary, stationary or nonstationary orientation of the magnetic vectors, only the number of {\it d} electrons and the magnitude of magnetic moments are modified during the self-consistency calculations. The orientation of magnetic vectors remains unaffected, i.e. spin rotations are completely decoupled from the self-consistency procedure in the NCAA model. This is different from DFT calculations, where the orientation of magnetic moments at a nonstationary point is modified during a self-consistency calculation unless local constraining fields holding predefined magnetization direction are introduced~\cite{stocks_98}.

The force acting on the orientation of the magnetic moments can be approximated using finite differences of the total energy evaluated for slightly different orientations. However, this is an inefficient approach as at least $2P+1$ self-consistent calculations need to be carried out for each 
state of  
a system containing
$P$ magnetic moments. 
A more efficient, direct method
for determining the force without additional self-consistency calculations is needed for large scale simulation of dynamics, optimization of transition paths, or, in general, navigation on the energy surface of a magnetic system. Analogous to the Hellmann-Feynman theorem~\cite{hellmann-feynman} of quantum mechanics, 
force theorems have been derived within DFT formalism for the gradient with respect 
to the position of atomic nuclei~\cite{andersen} and, more importantly for the present case, 
orientation of magnetic moments~\cite{liechtenstein_84,liechtenstein_87,oswald_85}.  
We present here a magnetic force theorem for the NCAA model which not only makes it possible to calculate the energy gradient without repeated self-consistency calculations, but also provides
a formula for the force acting on the orientation of the magnetic moments. 

The article is organized as follows. In the 
following
section, the NCAA model is briefly described as well as the method used in 
the 
self-consistency calculations. In Sec. III, the magnetic force theorem is derived and the results used to obtain a formula for magnetic forces. 
In Sec. IV,  the method is applied to a transition between parallel and antiparallel states in an Fe trimer and then 
to magnetization reversals in rectangular monolayer islands of Fe supported on W(110) surface. 
Finally, it is shown that a slightly different choice of the parameter values can lead to the appearance of a noncollinear metastable state with an antivortex structure~\cite{waeyenberge_06} in a supported island. 
Section V gives 
a 
summary.


\section{Noncollinear Alexander-Anderson model}

The AA model~\cite{alexander_64} 
extended
to multiple impurities and noncollinear ordering 
has
been described elsewhere~\cite{uzdin_98,uzdin_99}, 
but for completeness and to define the notation needed for the following sections, a summary of the most important equations is given here. 
In the AA model, the electronic structure of a 3{\it d} transition metal is approximated by two bands: one representing quasilocalized {\it d} electrons and the other representing itinerant {\it s}({\it p}) electrons. The Hamiltonian  
for a system of $P$ magnetic atoms is written as
\begin{equation}
\label{eq:AA}
\begin{split}
\EuScript{H}&=\sum\limits_{\bf{k},\alpha}\varepsilon_{\bf k}n_{\bf k\alpha}+ \sum\limits_{i,\alpha}\varepsilon_i^0n_{i\alpha}+ \sum\limits_{{\bf k},i,\alpha}\left(\upsilon_{i\bf k}d_{i\alpha}^{\dag}c_{\bf k\alpha}+\upsilon_{{\bf k} i}c_{\bf k\alpha}^{\dag}d_{i\alpha}\right)\\ &+\sum\limits_{i\ne j,\alpha}\upsilon_{ij}d_{i\alpha}^{\dag}d_{j\alpha}+ \frac{1}{2}\sum\limits_{i,\alpha} U_in_{i\alpha}n_{i-\alpha}.
\end{split}
\end{equation}
Here $d_{i\alpha}^{\dag}(d_{i\alpha})$ and $c_{\bf k\alpha}^{\dag}(c_{\bf k\alpha})$ are creation (annihilation) operators for {\it d} electrons localized on atom {\it i} and itinerant {\it s}({\it p})-electrons with momentum {\bf k}, respectively; 
$n_{i\alpha}=d_{i\alpha}^{\dag}d_{i\alpha}$, $n_{\bf k\alpha}=c_{\bf k\alpha}^{\dag}c_{\bf k\alpha}$ are corresponding occupation number operators. Greek indices denote spin projection ($\alpha,\beta=\pm$).
The energy of noninteracting {\it s}({\it p}) electrons, $\varepsilon_{\bf k}$, and {\it d} electrons, $\varepsilon_i^0$, hybridization parameters, $\upsilon_{i\bf k}$, hopping parameters, $\upsilon_{ij}$, and Coulomb repulsion between electrons with opposite spin projection, $U_i$, are spin independent. The last term in the Hamiltonian in Eq.~(\ref{eq:AA}), $U_in_{i\alpha}n_{i-\alpha}$, describes interaction between {\it d} electrons at atom $i$, $i=1,\ldots, P$. 

The Hamiltonian in Eq.~(\ref{eq:AA}) is invariant with respect to the choice of quantization axis. 
In order to describe noncollinear magnetic states, we will use a mean-field approximation at each site $i$ where a local quantization axis, $z_i$, is chosen to be along the local magnetic moment associated with atom $i$. 
The mean-field approximation for 
the
last term in the Hamiltonian is 
\begin{equation}
\label{eq:mean_field}
n_{i\alpha}n_{i-\alpha} \ \approx \   n_{i\alpha}\langle n_{i-\alpha}\rangle + \langle n_{i\alpha}\rangle n_{i-\alpha} - \langle n_{i\alpha}\rangle\langle n_{i-\alpha}\rangle,
\end{equation}
where $\langle n_{i\alpha}\rangle$ denotes the 
expectation value of an occupation number.
The mean-field approximation is invoked at each site $i$ for the electron operators $\tilde{d}_{i\alpha}^{\dag}$ and $\tilde{d}_{i\alpha}$ whose quantization axis is $z_i$. 
In the end, the mean-field Hamiltonian can be rewritten in terms of $d_{i\alpha}^{\dag}$ and $d_{i\alpha}$ where the
quantization axis is taken to be the laboratory $z$-axis,
the same for all sites $i$ 
(this procedure is described in detail in~\cite{hirai_92}, \S2, for a similar Hamiltonian). 

The {\it d} electrons are included explicitly here, while the influence of the itinerant {\it s}({\it p})-electrons is indirectly 
taken into account via the renormalization of model parameters. The mean-field Hamiltonian associated with
the {\it d} electrons, $H\equiv\EuScript{H}_{MF}^{(d)}$, is given by
\begin{equation}
\label{eq:AA_MF}
H=\sum\limits_{i,\alpha}E_{i}^\alpha n_{i\alpha}+\sum\limits_{i, j,\alpha,\beta}V_{ij}^{\alpha\beta}d_{i\alpha}^{\dag}d_{j\beta}- \frac{1}{4}\sum\limits_i U_i\left(N^2_i-M^2_i\right),
\end{equation}
where 
\begin{equation}
\label{eq:d_level}
E_i^{\alpha} = E_i^0 + \frac{U_i}{2}\left(N_i-\alpha \cos \theta_i M_i\right),
\end{equation}
\begin{equation}
\label{eq:d_hopp}
V_{ij}^{\alpha\beta} = \frac{U_i}{2}\left(\delta^{\alpha\beta}-1\right)\delta_{ij}\exp \left(-\alpha \mathrm i \phi_i\right)\sin \theta_i M_i + \left(1-\delta_{ij}\right)\delta^{\alpha\beta}V_{ij}.
\end{equation}
Here, $E_i^0$ is a renormalized 
energy
of unperturbed {\it d} levels and $V_{ij}$ are the hopping parameters that now contain both a contribution due to direct exchange between {\it d} states localized on sites $i$ and $j$ and a contribution from indirect {\it d}-{\it d} coupling through the conduction band. 
The choice of values for $E_i^0$ and $U_i$ 
depends mainly on the type of atom $i$, while the hopping parameters $V_{ij}$ 
also depend on the geometry of the system, in particular the distance between atoms $i$ and $j$. 
The hybridization of {\it s}({\it p}) and {\it d} bands leads to broadening of the {\it d} band
and the width, $\Gamma$, is assumed to be a parameter in the model.

The polar angle $\theta_i$ and the azimuthal angle $\phi_i$ define the direction of the $i$th magnetic moment
with respect to the laboratory quantization axis, $z$. 
In Eqs.~(\ref{eq:AA_MF})-(\ref{eq:d_hopp}), the number of {\it d} electrons, $N_i$, and the magnitude of magnetic moment, $M_i$, associated with
one of the  
five
degenerate {\it d} orbitals at atom $i$ are
\begin{equation}
\label{eq:N}
\begin{split}
N_i &= \langle \tilde{d}_{i+}^{\dag}\tilde{d}_{i+}\rangle + \langle \tilde{d}_{i-}^{\dag}\tilde{d}_{i-}\rangle,\\
M_i &= \langle \tilde{d}_{i+}^{\dag}\tilde{d}_{i+}\rangle - \langle \tilde{d}_{i-}^{\dag}\tilde{d}_{i-}\rangle
\end{split}
\end{equation}
For a given orientation of the magnetic moments, specified 
by the angles $\theta_i$ and $\phi_i$, the magnetic structure of a system of $P$ 
metal atoms is described by a set of self-consistent values of $N_i$ and $M_i$, 
where
$i = 1,\ldots,P$. 

The Green function $G(\omega)$, $G(\omega)=~\left[\omega -H\right]^{-1}$, is used to obtain the self-consistency condition for $N_i$ and $M_i$. According to Eqs.~(\ref{eq:N}):
\begin{equation}
\label{eq:occup_n}
N_i =\frac{1}{\pi}\int\limits_{-\infty}^0 d\omega\Im \left[G_{ii}^{++}(\omega- i \Gamma)+G_{ii}^{--}(\omega- i \Gamma)\right],
\end{equation}
\begin{equation}
\label{eq:occup_m}
\begin{split}
M_i &= \frac{1}{\pi}\int\limits_{-\infty}^0 d\omega\Im\left[G^{++}_{ii}(\omega- i \Gamma)-
G^{--}_{ii}(\omega- i \Gamma)\right]\cos{\theta_i} \\
&+\frac{1}{\pi}\int\limits_{-\infty}^0d\omega\Im\left[G^{+-}_{ii}(\omega- i \Gamma)e^{ i\phi_i}+
G^{-+}_{ii}(\omega- i \Gamma)e^{-i\phi_i}\right]\sin{\theta_i}
\end{split}
\end{equation}
Here $G^{\alpha\beta}_{ii}(\omega)$ denotes a matrix element of the Green function.
The magnetic system is assumed to be in contact with a large substrate which fixes the 
Fermi energy, for example a magnetic 
cluster or a thin film supported on a metal surface. 
The zero of energy is taken to be the Fermi energy ($\varepsilon_F = 0$). 

When self-consistency is achieved, the total energy of {\it d} electrons can be found as
\begin{equation}
\label{eq:self_en}
E = \frac{5}{\pi}
\int\limits_{-\infty}^0{d\omega \omega \Im{\Tr{G^*(\omega-\mathrm i \Gamma)}}} - 
5\sum_i{\frac{U_i}{4}\left(N_i^{*2}-M_i^{*2}\right)},
\end{equation}
where 
the
factor 5 is due to five-fold degeneracy of {\it d} orbitals.
Quantities marked with an asterisk correspond to  
self-consistent 
values.

An integration over the density of {\it d} states needs
to be carried out repeatedly 
in the self-consistency calculations
(see Eqs.~(\ref{eq:occup_n})-(\ref{eq:occup_m})). 
An efficient approach 
has been described in the literature~\cite{haydock_72,haydock_75,uzdin_98}. 
First of all, the recursion
method is applied in order to represent the Green function in terms of a continued fraction 
(see~\S3 in~\cite{haydock_72} and~\S2 in~\cite{haydock_75}).
In Appendix \ref{app_1}, we briefly describe how to obtain the continued fraction representation for the off-diagonal elements of the Green function. Then, the continued fraction is expanded in a series of partial fractions 
(see~\cite{uzdin_98} and also Appendix \ref{app_2} where the method is sketched), 
and a matrix element of the Green function then takes the form
\begin{equation}
\label{eq:g_sim_frac}
G_{ij}^{\alpha\beta}(\omega) = \sum_k \frac{p_k}{\omega - q_k},
\end{equation}
where the numbers $p_k$, $q_k$ depend on the orientation of the magnetic moments as well as on indices $i$, $j$ and $\alpha$, $\beta$. As a result, the density of states is expressed in terms of Lorentz profiles and can be integrated analytically. 

The total energy of the system 
can be
expressed analytically in terms of parameters of 
the self-consistent Hamiltonian
as
\begin{equation}
\label{eq:en_fast}
\begin{split}
E &= \frac{5}{\pi}\int\limits_{-\infty}^0 d\omega \omega \Gamma \sum_{\mu=1}^{2P} \frac{1}{\left(\omega - \omega^*_\mu\right)^2+\Gamma^2} - 5\sum_{i=1}^P \frac{U_i}{4}\left(N_i^{*2}-M_i^{*2}\right) \\
&= \frac{5}{\pi}\sum_{\mu=1}^{2P} \left[\omega^*_\mu\arcctg \frac{\omega^*_\mu}{\Gamma}+\frac{\Gamma}{2}\ln\left(\frac{\omega_\mu^{*2}}{\Gamma^2}+1\right)\right]- 5\sum_{i=1}^P \frac{U_i}{4}\left(N_i^{*2}-M_i^{*2}\right),
\end{split}
\end{equation}
where $\omega^*_\mu$ are the eigenvalues of $H^*$. 

In self-consistency procedure, where the number of {\it d} electrons and magnitude of magnetic moments are found for an arbitrary orientation of magnetic vectors,
we use a fundamental assumption about the
hierarchy of relaxation time scales. Relaxation of the diagonal components of the spin density matrix, which in a local frame of reference give the number of {\it d} electrons and magnitude of magnetic moments, is much faster than the relaxation
of the off-diagonal components which give the orientation 
of the magnetic moments~\cite{antropov_96}. Thus, $N$ and $M$ are treated as fast degrees of freedom
which adjust instantaneously to the orientation of magnetic moments defined by polar and azimuthal angles 
$\bm{\theta}$ and $\bm{\phi}$, the slow degrees of freedom. This is analogous to the Born-Oppenheimer approximation in atomic systems where the electronic degrees of freedom are
assumed to be fast as compared to slowly varying positions of nuclei, 
and the total energy of the system is
expressed as
a function of the slow degrees of freedom only. 


\section{Magnetic force theorem}

As discussed in the introduction, various calculation require the evaluation of the gradient of the 
total energy of the system which in the present case is the force 
acting on the orientation of the magnetic vectors. 
Below, we present a magnetic force theorem for the NCAA model which makes it possible to express the force
in terms of self-consistent values of the number of {\it d} electrons,$N^*$, and modulus of the magnetic moment, $M^*$. 
The theorem is equivalent to a variational principle according to which self-consistency corresponds to a stationary point of the energy as a function of the fast degrees of freedom, $N$ and $M$:
\begin{align}
\forall i \colon \quad \left.\frac{\partial E}{\partial N_i}\right|_{\substack{N=N*\\M=M*}}&=0 
\ \ \ \  {\rm and} \ \ \ 
\quad \left.\frac{\partial E}{\partial M_i}\right|_{\substack{N=N*\\M=M*}} =0.
\label{eq:theor1}
\end{align}

We will need two lemmas for the Green function that are proved in Appendix C:
\begin{equation}
\label{eq:lemma1}
\frac{\partial\Tr{G(\omega)}}{\partial N_i} = -\frac{U_i}{2}\frac{\partial}{\partial \omega}\left(G_{ii}^{++}(\omega)+G_{ii}^{--}(\omega)\right),
\end{equation}
and
\begin{equation}
\label{eq:lemma2}
\begin{split}
\frac{\partial\Tr{G(\omega)}}{\partial M_i} &= \frac{U_i}{2}\frac{\partial}{\partial \omega}\left[\left(G_{ii}^{++}(\omega)-G_{ii}^{--}(\omega)\right)\cos{\theta_i}\right.\\
&+\left.\left(G_{ii}^{+-}(\omega)e^{i\phi_i}+G_{ii}^{-+}(\omega)e^{-i\phi_i}\right)\sin{\theta_i}\right].
\end{split}
\end{equation}

According to Eqs. ~(\ref{eq:self_en}) and~(\ref{eq:lemma1}) 
\begin{align*}
\frac{\partial E}{\partial N_i} &= \frac{5}{\pi}
\int\limits_{-\infty}^0{d\omega \omega \Im{\frac{\partial}{\partial N_i}\Tr{G(\omega-\mathrm i \Gamma)}}} - 
5\frac{U_i}{2}N_i\\
&=5\frac{U_i}{2}\left[\frac{1}{\pi}
\int\limits_{-\infty}^0{d\omega \Im{\left(G_{ii}^{++}(\omega-\mathrm i \Gamma)+G_{ii}^{--}(\omega-\mathrm i \Gamma)\right)}} - N_i\right],
\end{align*}
where integration by parts is invoked. According to Eq.~(\ref{eq:occup_n}), 
the expression in the square brackets is equal to zero when self-consistency 
has been
reached. 

The equation for the derivative with respect to $M_i$ in Eq.~(\ref{eq:theor1}) 
is proved in the same way. Using~(\ref{eq:self_en}) and~(\ref{eq:lemma2}), we obtain
\begin{align*}
\frac{\partial E}{\partial M_i} =& \frac{5}{\pi}\int\limits_{-\infty}^0{d\omega \omega \Im{\frac{\partial}{\partial M_i}\Tr{G(\omega-\mathrm i \Gamma)}}} + 
5\frac{U_i}{2}M_i\\
=&5\frac{U_i}{2}\Biggl\{-\frac{1}{\pi}
\int\limits_{-\infty}^0 d\omega \Im\left[\left(G_{ii}^{++}(\omega-\mathrm i \Gamma)-G_{ii}^{--}(\omega-\mathrm i \Gamma)\right)\cos{\theta_i}\right.\Biggr.\\
&+
\Biggl.\left.\left(G_{ii}^{+-}(\omega-\mathrm i \Gamma)e^{i\phi_i}+G_{ii}^{-+}(\omega-\mathrm i \Gamma)e^{-i\phi_i}\right)\sin{\theta_i}\right] + M_i\Biggr\}.
\end{align*}
Due to Eq.~(\ref{eq:occup_m}), 
the expression in the curly brackets is equal to zero when $M=M^*$.

The magnetic force theorem can be used to derive an expression for the force acting
on the orientation of the magnetic moments within the NCAA model. 
According to the force theorem, a
derivative of the energy, $E=E(\lambda)$, with respect to a parameter $\lambda$ (a slow degree of freedom) can be computed from the explicit dependence only, without having to include implicit dependence
\begin{equation}
\label{eq:force2}
\frac{d E(\lambda)}{d \lambda} = \frac{\partial E(\lambda)}{\partial \lambda} = \frac{5}{\pi}
\int\limits_{-\infty}^0{d\omega \omega \Im{\Tr{\frac{\partial G^*(\omega-\mathrm i \Gamma;\lambda)}{\partial \lambda}}}}.
\end{equation}
Here, 
${\partial G^*(\omega-\mathrm i \Gamma;\lambda)} / {\partial \lambda}$ 
can be found by using the resolvent identity
\begin{equation}
\label{eq:diff_green}
\frac{\partial G(\omega;\lambda)}{\partial \lambda} = G(\omega;\lambda)\frac{\partial H(\lambda)}{\partial \lambda}G(\omega;\lambda),
\end{equation}
which, together with Eq.~(\ref{eq:force2}), gives
\begin{equation}
\label{eq:force3}
\begin{split}
\frac{d E(\lambda)}{d \lambda} &= \frac{5}{\pi}
\int\limits_{-\infty}^0{d\omega \omega \Im{\Tr{\left[G^*(\omega-\mathrm i \Gamma;\lambda)\frac{\partial H^*(\lambda)}{\partial \lambda}G^*(\omega-\mathrm i \Gamma;\lambda)\right]}}}\\
&=\frac{5}{\pi}
\int\limits_{-\infty}^0{d\omega \Im{\Tr{\left[G^*(\omega-\mathrm i \Gamma;\lambda)\frac{\partial H^*(\lambda)}{\partial \lambda}\right]}}},
\end{split}
\end{equation}
that is, the derivative of the total energy with respect to a parameter coincides with the expectation value of the derivative of the Hamiltonian with respect to that parameter, 
analogous to the Hellmann-Feynman theorem~\cite{hellmann-feynman}. 

In practice, it is convenient to calculate the trace in Eq.~(\ref{eq:force3}) using the basis in which $H^*(\lambda)$ and $G^*(\omega;\lambda)$ are diagonal
\begin{equation}
\label{eq:trace}
\Tr{\left[G^*(\omega;\lambda)\frac{\partial H^*(\lambda)}{\partial \lambda}\right]}=\sum_{\mu=1}^{2P}\frac{\xi_\mu^*}{\left(\omega-\omega_\mu^*\right)},
\end{equation}
where $\xi_\mu^*$ are the diagonal elements of 
${\partial H^*(\lambda)} / {\partial \lambda}$ in the relevant basis.
The integral in Eq.~(\ref{eq:force3}) can then be evaluated analytically leading to
\begin{equation}
\label{eq:force4}
\frac{d E(\lambda)}{d\lambda} = \frac{5}{\pi}\sum_{\mu=1}^{2P}\xi_\mu^*\arcctg\left(\frac{\omega_\mu^*}{\Gamma}\right).
\end{equation}
With $\lambda=\theta_i$ or $\lambda=\phi_i$ and $i = 1,\ldots,P$, this gives the gradient of the energy 
with respect to the angles defining the orientation of the magnetic moments. 

The procedure for evaluating the energy gradient is as follows:
First, derivatives of the self-consistent Hamiltonian,  
${\partial H^*(\bm{\theta},\bm{\phi})}/{\partial \theta_i}$ 
and  
${\partial H^*(\bm{\theta},\bm{\phi})} / {\partial \phi_i}$, 
which are given explicitly by 
\begin{align}
\left(\frac{\partial H^*(\bm{\theta},\bm{\phi})}{\partial \theta_i}\right)_{kj}^{\alpha\beta} &= \frac{1}{2}\delta_{ji}\delta_{ki}U_i M_i^*\left[\alpha\delta^{\alpha\beta} \sin \theta_i+
\left(\delta^{\alpha\beta}-1\right)\exp \left(-\alpha \mathrm i \phi_i\right)\cos \theta_i\right],\\
\left(\frac{\partial H^*(\bm{\theta},\bm{\phi})}{\partial \phi_i}\right)_{kj}^{\alpha\beta} &=\frac{\mathrm i \alpha}{2}\left(1-\delta^{\alpha\beta}\right)\delta_{ji}\delta_{ki}U_i M_i^* e^{-\alpha \mathrm i \phi_i} \sin \theta_i
\end{align}
are transformed to a basis where $H^*(\bm{\theta},\bm{\phi})$ is diagonal. Then, their diagonal matrix elements, $\xi_\mu^*(\theta_i)$, $\xi_\mu^*(\phi_i)$, are inserted into Eq.~(\ref{eq:force4}) and the derivatives 
with respect to $\theta$ and $\phi$ evaluated.

Thus, after self-consistency has been reached for an arbitrary, in general nonstationary, orientation of magnetic vectors, magnetic forces are readily available; no additional self-consistency calculations need to be performed. The forces can then be used to guide the orientation of magnetic vectors in spin dynamics simulations, minimization of the energy, or calculations of MEPs.

\section{Applications}

In what follows, we will demonstrate how the 
NCAA model and magnetic force theorem can be used to find 
(meta)stable magnetic states as well as MEPs for transitions between these states.
Given some, possibly random initial values of the angles specifying the orientation of the magnetic moments, 
a steepest descent or, more efficiently, a conjugate gradient minimization of the energy can be used to find a configuration corresponding to a minimum on the energy surface, and thus a stable or metastable magnetic state. 
The ability to 
evaluate the gradient of the energy with respect to the angles specifying the orientation of the magnetic moments makes
such calculations fast and reliable and improves the chances of finding novel and unexpected magnetic states. This is 
illustrated by an example below. 
 
In order to assess the thermal stability of a magnetic state and to estimate the rate of transitions to other states,
it is useful to find MEPs.
An MEP shows how each magnetic moment rotates during the transition
in an optimal way so as to make the energy barrier minimal. 
Thus, MEPs play a key role in the rate theory for magnetic transitions as they represent paths of highest statistical 
weight and reveal the transition mechanism and activation energy.
The nudged elastic band (NEB) method can be used to find MEPs~\cite{NEBleri}.
There, an initial path represented by a chain of intermediate states, or 'images', 
which give a discrete representation of the path,
is created between a pair of stable states. An iterative algorithm involving force projections and a minimization algorithm is then used to bring the images to the nearest MEP. Each image is defined by a point in configuration space, i.e. by a set of angles $\bm{\theta}$ and $\bm{\phi}$ for each magnetic moment. In order to ensure continuity of the path and control the distribution of the images along the path, springs are introduced between adjacent images. 
At each step in the iteration, 
an estimate of the local tangent to the path at each image is made and the
images moved only according to the perpendicular component of the force. 
In order to distribute the images in a predefined way along the path, for example equally, a spring force is included
between the images but only the component parallel to the path is included when the images are moved ~\cite{NEBleri}.
After convergence, when the projected forces are zero, the images give a discrete representation of the MEP. 
As the forces need to be evaluated repeatedly during this optimization procedure, the analytical expression based on 
the magnetic force theorem is of great 
importance.


\subsection{Fe trimer on a metallic substrate}

Previous studies have shown that ad-trimers of Fe, Cr and Mn can have several magnetic states that 
are close in energy~\cite{uzdin_01}. 
We use 
a trimer to
illustrate
the methodology presented here
because the energy surface with two magnetic states and the MEP 
connecting
them can be visualized easily. 
The parameters in the NCAA model were chosen to be representative of an Fe trimer on a metal surface 
(see  ref.~\cite{uzdin_12} and references therein): 
$E^0/\Gamma=-12$ and $U/\Gamma=13$. 
These values are only 5\% different from values used to model bulk iron~\cite{uzdin_12,uzdin_09a,uzdin_09}.
Here we have chosen  
unequal
values for the three hopping parameters
$V_{12}/\Gamma=1.0$, $V_{13}/\Gamma=1.19$ and $V_{23}/\Gamma=1.22$ 
corresponding to an asymmetric geometry of the trimer. 

\begin{figure}
\begin{center}
\includegraphics[scale=0.3]{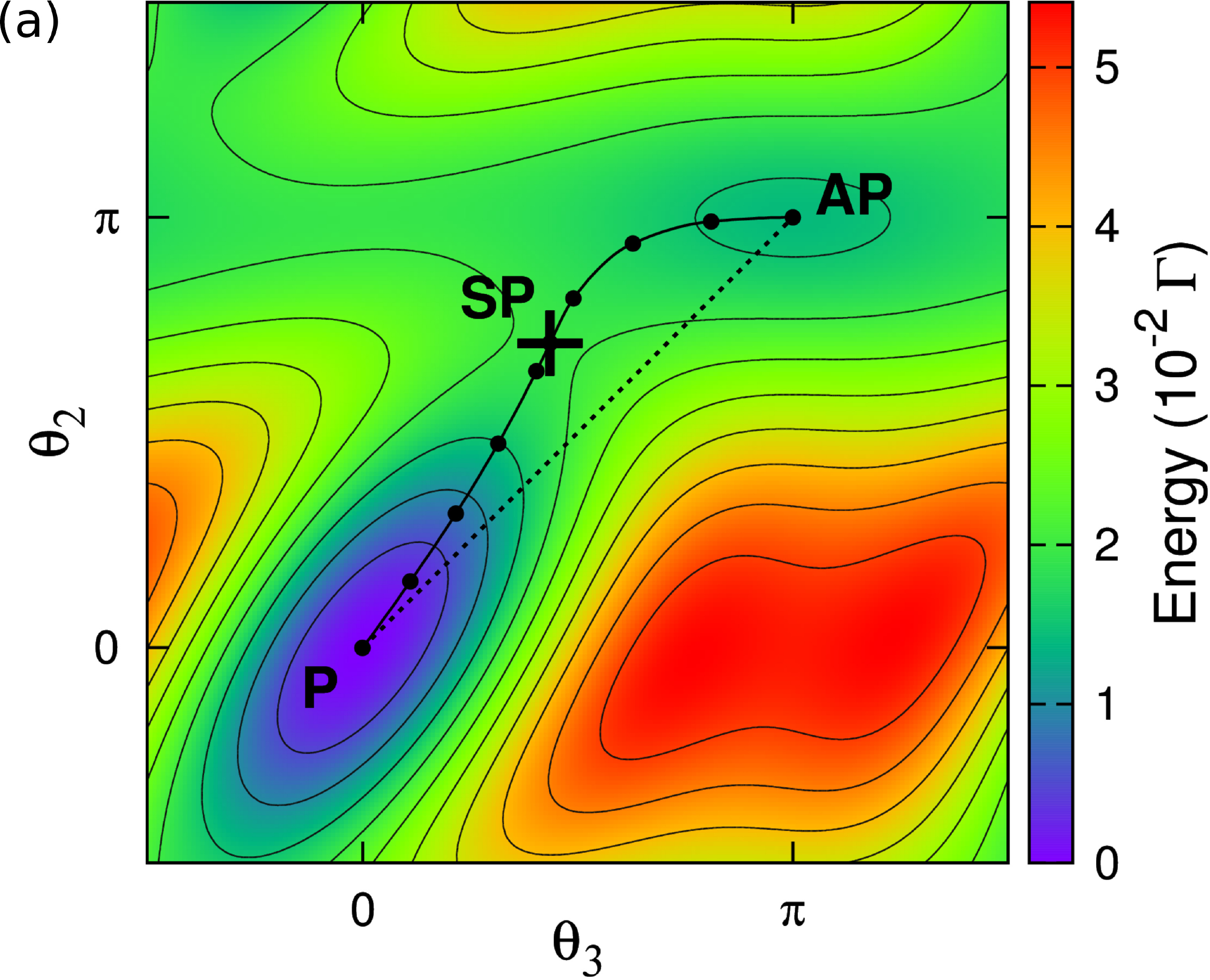}\ \ \ \ \ \ \includegraphics[scale=0.3]{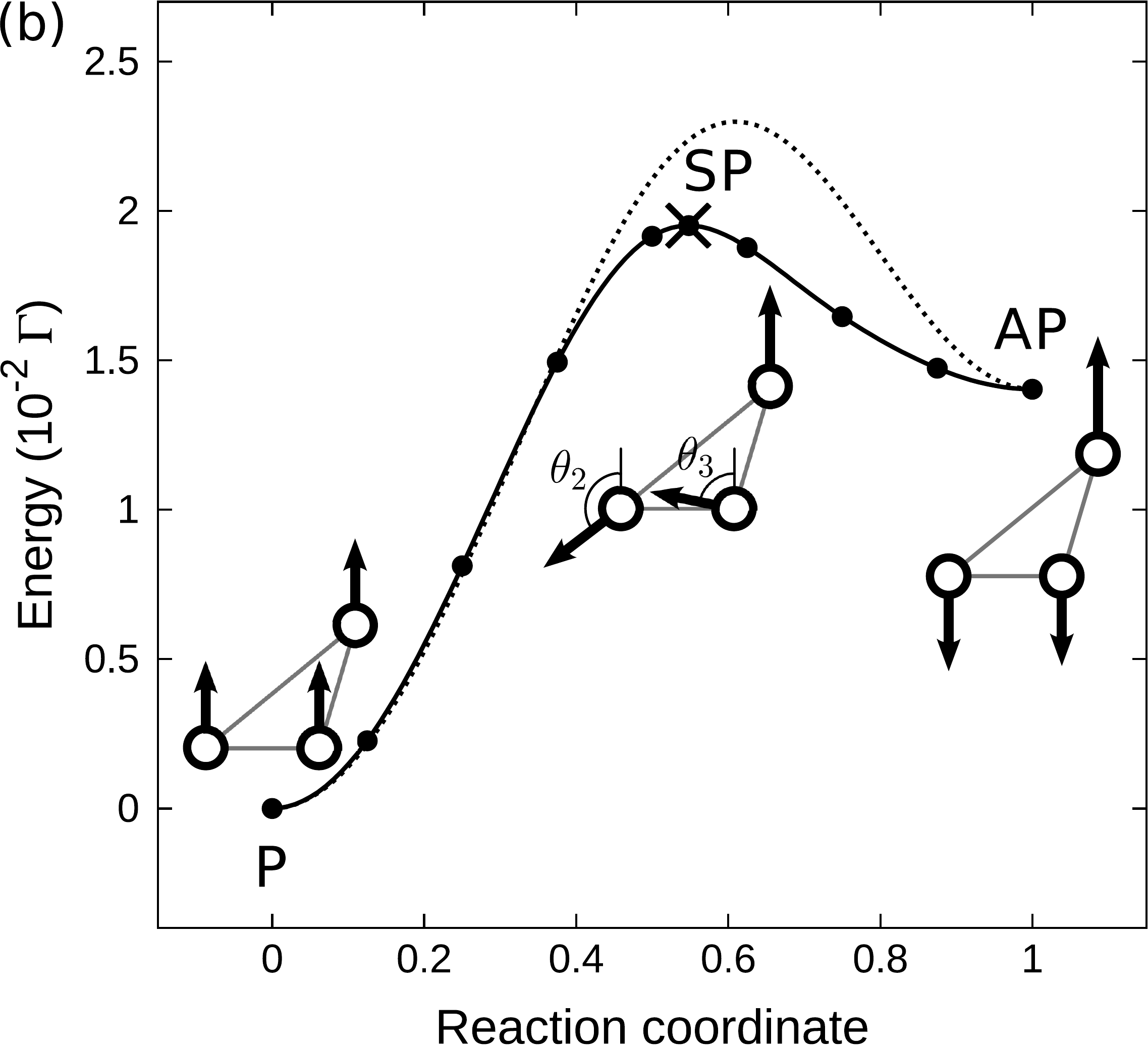}
\caption{
The minimum energy path between two local minima on the energy surface corresponding to parallel (P) and antiparallel (AP) ordering of the spins is shown with a solid line. The filled circles show the location of images in a nudged
elastic band calculation.  $\times$ shows the location of the first order saddle point. 
The dotted line corresponds to the uniform rotation path.
(a)  Energy surface for an Fe trimer evaluated with the NCAA model.  
(b)  Energy along the minimum energy and uniform paths. 
Insets show the direction and the magnitude of the magnetic moments at 
the minima and at the saddle point. The middle inset shows the definition of the angles $\theta_2$ and $\theta_3$.
The reaction coordinate is defined as the sum of rotations of all magnetic moments along the path normalized
by 
its
total length.
}
\label{fig:Trimer} 
\end{center}
\end{figure}
%
Spin-orbit interaction is not taken into account in the NCAA model. The total energy of the system is, therefore, invariant under the uniform rotation of all the magnetic moments and only relative orientation of magnetic moments is relevant. Moreover, it is possible to show that the magnetic moments of all the atoms will tend to lie in a plane. 
The energy of the system increases significantly when one of the magnetic moments of an atom points out of a plane
formed by the other two. 
Therefore, it is sufficient to set $\phi_i=0$ for all three atoms to visualize the relevant part of the energy surface. 
It is convenient to choose the quantization axis for the system to be along the magnetic moment of one of the atoms,
and  
a configuration 
of the system
is then 
fully
specified by only two angles, $\theta_2$ and $\theta_3$
between magnetic moments of the first and second atom and 
between
first and third atom, respectively 
(see inset in Fig.~\ref{fig:Trimer}(b)).

Fig.~\ref{fig:Trimer}(a) shows a
contour graph of the energy surface, $E(\theta_2,\theta_3)$. 
It has
two minima, one at $\theta_2=\theta_3=0$ and 
the other at $\theta_2=\theta_3=\pi$, and a maximum at $\theta_2=0$, $\theta_3=\pi$. Both minima correspond to collinear spin states of the trimer. 
The lower energy state, $P$, has all magnetic moments pointing in the same direction, 
but the metastable state, $AP$, has the magnetic moment of the first atom pointing in the opposite direction to 
the other two.
This is an example of a magnetic system with two possible states corresponding to two different arrangements of the magnetic vectors. 

In the NEB calculations of the MEP, the initial path was chosen to correspond to uniform rotation.
After convergence, the path lies through the first order saddle point on the energy surface and has a lower maximum 
than the uniform rotation path, see  Fig.~\ref{fig:Trimer}.
Note that not only orientation of magnetic moments changes during the transition but also their magnitudes (see the  insets in Fig.~\ref{fig:Trimer}(b)).

Fig.~\ref{fig:Trimer}(b) shows the variation in the total energy along the MEP. The energy maximum along the MEP corresponds to the first order saddle point on the energy surface. 
It
gives an estimate of the activation energy barrier for magnetic transitions within harmonic transition state theory \cite{bessarab_12}. 
The barrier for the transition $P\rightarrow AP$ was found to be $E_{AP\leftarrow P}=E^S-E^P=0.019$~$\Gamma$, while for the reverse transition it is smaller, $E_{P\leftarrow AP}=E^S-E^{AP}=0.005$~$\Gamma$. Energy variation for the uniform rotation of magnetic moments is also shown in Fig.~\ref{fig:Trimer}(b) for comparison. 
This simple example illustrates the methodology 
which is applied to
larger and more complex systems below.


\subsection{Fe on W(110)}

Recently, extensive experimental data on thermally induced magnetization reversals has been 
reported for monolayer Fe islands on W(110) surface~\cite{krause_09}.
We have,
for comparison,
 used the methodology presented above to calculate the magnetism and rate of transitions in 
rectangular Fe islands
of varying shape and size.
Previously, a theoretical analysis using a Heisenberg-type Hamiltonian 
had been
carried out~\cite{bessarab_13}, 
but here we present results using the NCAA model.
A brief account of these calculations has been given elsewhere~\cite{bessarab_13a}. 

Since the NCAA model does not include spin-orbit coupling, 
it is necessary to supplement it with additional terms which introduce magnetic anisotropy. 
These terms describe the interaction of the magnetic system, here the Fe island, with the substrate and makes 
the magnetic vectors lie preferably along a particular direction within the surface plane.
The total energy of the system is then
\begin{equation}
E = E^{NCAA} + \sum_{n}{K_n}\sum_i{(\bm{M}_i\cdot \bm{e}_n)^2},
\label{eq:app4_en}
\end{equation}
where  $E^{NCAA}$ is given by Eq.~(\ref{eq:en_fast}). 
The index $n$ in the sum takes two values, $\parallel$ for easy-axis and $\perp$ for easy-plane anisotropy
representing the interaction with the substrate.   
An easy-plane anisotropy $K_{\perp}$ is included to make it preferable for the magnetic moments to lie in the (110) plane.
An easy-axis anisotropy $K_{\parallel}$ along the $[1\bar{1}0]$ direction is also included  
in order to give
the system 
two degenerate magnetic states, 
with
magnetic vectors aligned parallel or antiparallel to this axis. 
The parameter values were chosen to be 
$K_{\perp} \mu_B^2 = 0.7$ {\it meV} and $K_{\parallel} \mu_B^2 = -0.3$ meV.  A wide range of values 
for $K_{\perp}$ suffices to keep the magnetic moments within the 
surface plane, but the value of $K_{\parallel}$ was chosen to get roughly the  
experimentally determined
magnitude for the activation energy of
magnetization reversals in Fe islands on W(110).

Although magnetostatic dipole-dipole interaction was included in our previous study using a Heisenberg-type model~\cite{bessarab_13}, its contribution to the activation energy of the magnetization reversal was less than 0.5\%. The size of the Fe islands studied here is less than 6 nanometers. At this length scale, dipole-dipole interaction is irrelevant and, therefore, not included here.

The parameters $E^0/\Gamma$ and $U/\Gamma$ in the NCAA model were chosen to have the same values as
for the trimer, but the $V/\Gamma$ and $\Gamma$ parameters were chosen to reproduce results of DFT
calculations of an Fe monolayer on W(110), as described below.

\subsubsection{Fe monolayer on W(110)}

\begin{figure}[h!]
\centering
\includegraphics[width=0.7\columnwidth]{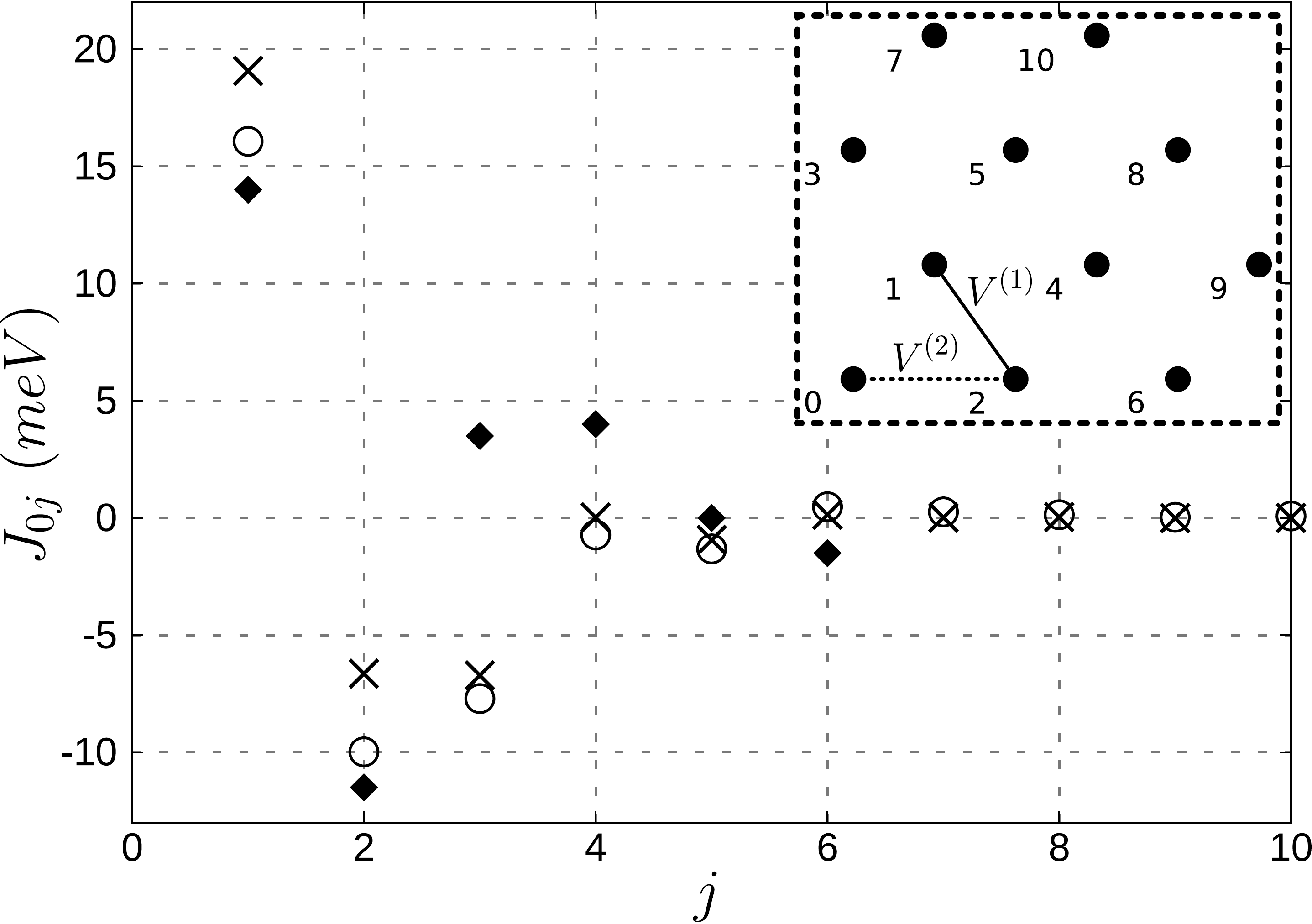}
\caption{
Exchange interaction parameters for a monolayer of Fe-atoms on W(110) in the collinear ground state. 
(Crosses) show results obtained when only nearest-neighbor hopping parameter is nonzero,
$V^{(1)}/\Gamma=0.9$ 
(circles) show results obtained when also second nearest-neighbor hopping is included,
$V^{(2)}/\Gamma=0.3$.
(Diamonds) show results of density functional theory calculations presented in ref.~\cite{bergman_10}. 
The ferromagnetic coupling between nearest neighbors and antiferromagnetic coupling between second-nearest
neighbors is nicely reproduced in the NCAA model, even when 
only nearest-neighbor hopping parameters are included.
The inset shows the position of neighbors $j$ relative to atom $0$ and illustrates the definition of the hopping parameters. 
}
\label{fig27}
\end{figure}

An Fe
monolayer
on the W(110) surface is commensurate with the substrate.
Each Fe atom in the layer has four nearest neighbors at a distance of 0.87$a$, where $a$ is the lattice constant of 
the W crystal and two second nearest neighbors at a slightly larger distance, $a$ (see the inset in Fig.~\ref{fig27}). 
The 
hopping parameter corresponding to the first nearest neighbors
was chosen to have
the value $V^{(1)}/\Gamma=0.9$ where $\Gamma = 0.2$ eV. 
For simplicity, the hopping parameters for further neighbors were set to zero.
This value of $\Gamma$ agrees well with the width of the {\it d}-band subpeaks in the density of states calculated 
with DFT, and with this value of $V^{(1)}/\Gamma$, the calculated 
magnetic moment is $2.4 \mu_B$ reproducing the results of 
DFT calculations~\cite{costa_08}.
Even though only nearest neighbor hopping parameters are 
non-zero, there is still exchange coupling between second and farther neighbors in the NCAA model because each pair of moments interacts due to indirect coupling through intermediate atoms. In order to demonstrate this, we calculated the exchange parameters, $J_{0j}$, 
for a collinear ground state, where indices $0$ and $j$ label atoms as illustrated in the inset in Fig.~\ref{fig27}. 
If the polar axis is chosen to be perpendicular to the (110) surface, then the exchange parameters are defined as 
$$
J_{0j} = -\frac{\partial^2 E}{\partial \phi_0 \partial \phi_j}
$$
and were evaluated by finite differences of the calculated gradients after small rotations.

The exchange coupling calculated for a monolayer of Fe on a W(110) surface using the NCAA model is in good agreement with results of DFT calculations~\cite{bergman_10}. 
In both cases, ferromagnetic exchange coupling is obtained between first nearest neighbors, while antiferromagnetic 
exchange coupling is 
obtained between the second nearest neighbors (see Fig.~\ref{fig27} and Fig.~1 in ref.~\cite{bergman_10}). 
For more distant neighbors,
this parametrization of the NCAA model, however, gives $J_{03} = J_{02}$, 
while the DFT results show a significant difference between the two. 
If additional parameters are included in the NCAA model, for example $V^{(2)}/\Gamma = 0.3$ for direct hopping 
between second nearest 
neighbors, then the values of $J_{02}$ and $J_{03}$ become different as shown in Fig.~\ref{fig27}
and agreement with DFT results for $J_{01}$ and $J_{02}$ is improved.
In what follows, we will, however, use the simplest possible parametrization of the NCAA and include only the hopping 
parameter between nearest neighbors, $V^{(1)}/\Gamma=0.9$.  This gives good agreement with the DFT results 
for the magnitude of the magnetic moment as well as the most important exchange coupling, 
between nearest neighbors and 
between
second nearest neighbors.  More elaborate parametrization of the NCAA model could be undertaken but will not be
pursued here.


\subsubsection{Fe islands on W(110)}

Calculations using the  
NCAA model with the parameters described above 
were carried out for monolayer, 
rectangular 
islands of Fe-atoms 
of varying shape and size.  As an example, the
magnetic moments 
obtained from the self-consistent calculations of a 29$\times$5 atomic row island are shown in Fig.~\ref{fig29}.  
The value obtained for the innermost atoms is nearly the same as for the full monolayer, but the atoms
at the rim of the island have about 10\% larger magnetic moment. In between, the atoms have a slightly smaller 
value than atoms in a full monolayer. 
The increased magnetic moment at the rim atoms can
explain in part island size dependence of the activation energy for magnetization reversals, as discussed below.

\subsubsection{Magnetization reversal}

\begin{figure}[h!]
\centering
\includegraphics[width=0.6\columnwidth]{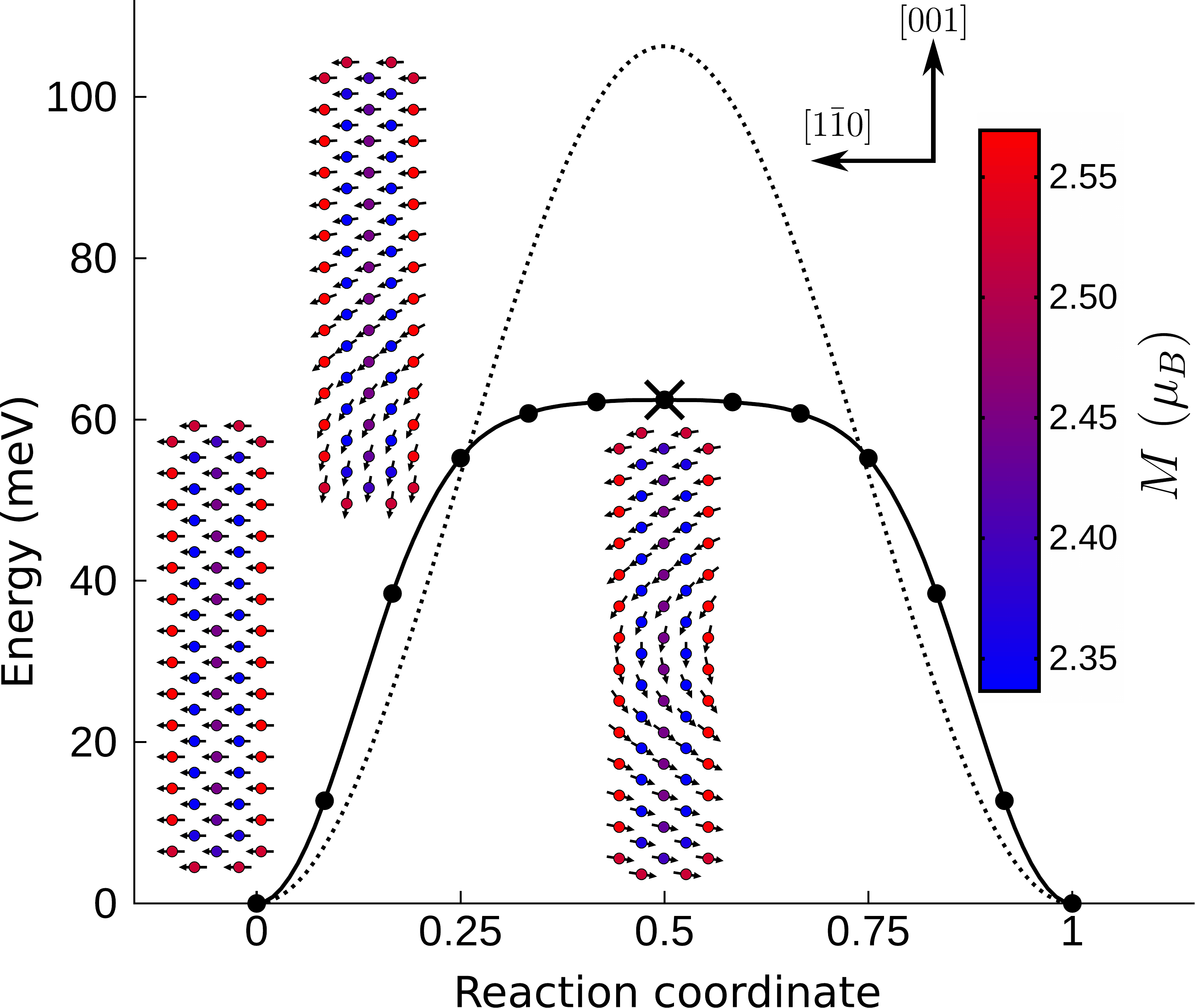}
\caption{
Magnitude of the magnetic moment of Fe-atoms in a $29\times 5$ atomic row island on W(110) surface, calculated
using the NCAA model, and energy change along the MEP (solid line) for a magnetization reversal. 
The energy maximum along the MEP, which corresponds to a saddle point on the energy surface, is marked with a 
$\times$. The energy change for a uniform rotation of magnetic moments is also shown (dotted line). The reaction coordinate is defined as the normalized displacement along the path. The orientation of crystallographic axes, $\left[1\bar{1}0\right]$ and $[001]$, are indicated with arrows. The anisotropy axis, $K_\parallel$, is oriented along the $\left[1\bar{1}0\right]$ direction. Insets show the direction and 
magnitude of the magnetic moments at the energy minimum, at the saddle point and at 
another intermediate configuration.
}
\label{fig29}
\end{figure}

There are two degenerate magnetic states of the islands, 
where all the magnetic vectors point in one of the two directions
along the anisotropy axis.
Thermally induced magnetization reversal transitions between these two states were studied by 
calculating MEPs, as described 
for the trimer
above.
The orientation of the magnetic moment of each atom was included explicitly. Two mechanisms for magnetization reversal were found. Small islands, with fewer than 15 atomic rows, reverse their magnetization by coherent rotation of all the magnetic moments. 
However, transitions in islands with more than 15 atomic rows along either side follow a more complicated path involving 
nucleation and propagation of an excitation
that can be described as a thin, temporary domain wall. This is similar to what we previously found in calculations using a Heisenberg-type Hamiltonian~\cite{bessarab_13} and what has been seen in atomistic spin dynamics simulations of similar systems~\cite{boettcher_11,bauer_11}.
Fig.~\ref{fig29} shows results for a $29\times 5$ atomic-row island which contains 72 Fe-atoms.
It turns out that the
variation in the magnitude of the magnetic moments along the MEP calculated self-consistently with the NCAA model  
is small, lending support for the application of a Heisenberg-type Hamiltonian to study such transitions.
The magnetization reversal starts at one of the narrower ends of the island and a domain wall forms 
parallel to the short
axis of the island. The domain wall then moves along the [001] direction eventually leading to reversal of the magnetization of the whole island. Fig.~\ref{fig29} shows the energy along the MEP. 
The height of the energy barrier is determined by the domain wall length which in this case scales with the size of the island along 
the $[1\bar{1}0]$ direction.
The energy change for a uniform rotation of magnetic moments is also shown in Fig.~\ref{fig29} for comparison.
This
shows how much the formation of the transient domain wall lowers the activation energy of the transition. 

\begin{figure}[h!]
\centering
\includegraphics[width=0.5\columnwidth]{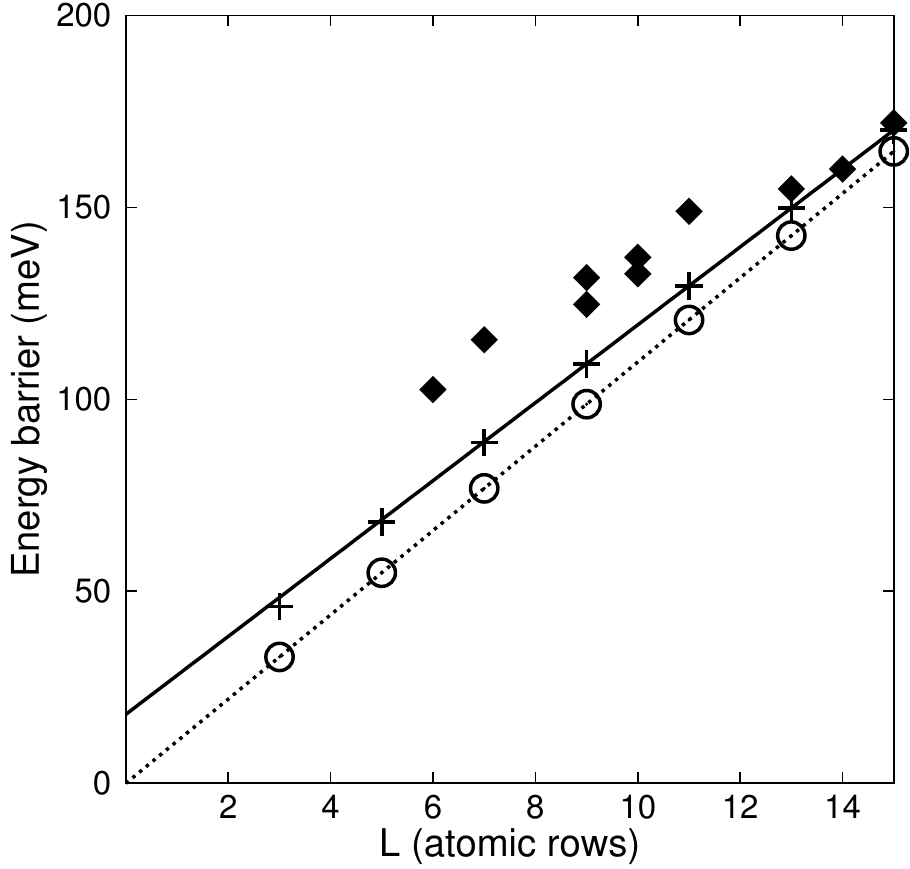}
\caption{
Energy barrier to magnetization reversal in rectangular islands of Fe-atoms on W(110) surface as a function of the island size, $L$, along the $[1\bar{1}0]$ direction 
calculated using
the NCAA model (crosses) and Heisenberg-type Hamiltonian (circles; data taken from Ref.~\cite{bessarab_13}). The length of the islands along the
[001] direction is 27 atom-rows in all cases. Solid and dotted lines represent linear fits, giving an extrapolation to islands with zero width, i.e. the intercept. 
Diamonds show experimentally measured energy barriers for islands which have a more irregular shape, 
see ref.~\cite{krause_09}.
}
\label{fig_rim}
\end{figure}

The ca. 10\% 
larger
magnetic moment of the rim atoms as compared with the innermost atoms does, however, 
lead to some difference from the results obtained
using a Heisenberg-type Hamiltonian~\cite{bessarab_13}, where the magnitude of the magnetic moments is 
taken
to be the same for all atoms.
Fig.~\ref{fig_rim} compares the calculated energy barrier as a function of the size of the islands, $L$, along the 
$[1\bar{1}0]$ direction obtained using the Heisenberg-type Hamiltonian (results are taken from Ref.~\cite{bessarab_13}) and the NCAA model. The number of atomic rows along the [001] direction equals 27 and is kept constant. For a vanishing domain wall length (the intercept with the y-axis), the Heisenberg-type model predicts zero activation energy barrier, but the NCAA model gives an offset of about 18 meV. 
This offset 
appears because the rim atoms have larger anisotropy energy due to
the larger magnetic moment.
As the islands  
become smaller,
the relative number of rim atoms increases. This leads to a larger activation energy for the small islands as compared to what the Heisenberg-type model predicts, where the anisotropy is the same for all atoms in the island. 

Such an offset has, in fact, been observed 
experimentally~\cite{krause_09}, see Fig.~\ref{fig_rim}, lending support for the NCAA model results.
The offset obtained from the measurements is, however, about three times larger than the one obtained from
our simple model.
Several effects not included in the calculations could contribute to this difference, such as
(1) irregular shape of the islands measured experimentally where larger number of atoms are at the rim
than in the rectangular islands of the calculation,
(2) broken symmetry in the spin-orbit interaction at the edges of islands leading to larger anisotropy, or
(3) impurity adatoms sitting preferably either on rim atoms or 
on interior atoms (but not both)~\cite{bessarab_13b}.


\subsubsection{An antivortex metastable state}

The analytical forces provided by the force theorem make it easier to navigate on the energy surface to find
local minima, corresponding to (meta)stable magnetic states with possibly complex, non-collinear ordering of magnetic 
moments.
Starting from a random initial 
orientation of the magnetic moments, a steepest descent or conjugate gradient minimization of the energy will bring the system to a (meta)stable state.
This procedure reveals metastable states which 
could
be hard to find otherwise.
We demonstrate this with a small island 
having
7x7 atomic rows.
Two of the parameters in the NCAA model have here been changed slightly: 
The energy of the {\it d} level with respect to the Fermi energy was changed from $E^0/\Gamma=-12$ to $-11.9$, and the hopping parameter between nearest neighbors was changed from $V^{(1)}/\Gamma=0.9$ to $1.025$. 
Such a slight change in the model parameters could be the result of an external perturbation such as an external 
electrical field or the presence of impurities or defects~\cite{bessarab_13b}. 
It does not lead to significant changes in the magnetic 
moments and exchange coupling. For 
the
Fe
monolayer
on W(110) in the collinear ferromagnetic state, 
the
magnetic moment increases by
only $0.01\mu_B$.
Also, the exchange interaction parameters $J_{0j}$ do not 
change much,
except for $J_{02}$ and $J_{03}$, which change from $-6.7$ $meV$ to $-10$ $meV$. However, this slight change in $E^0/\Gamma$ and $V^{(1)}/\Gamma$ is 
large
enough 
to have a significant effect on the magnetic structure of 
Fe 
nanoislands on W(110).

\begin{figure}[h!]
\centering
\includegraphics[width=0.7\columnwidth]{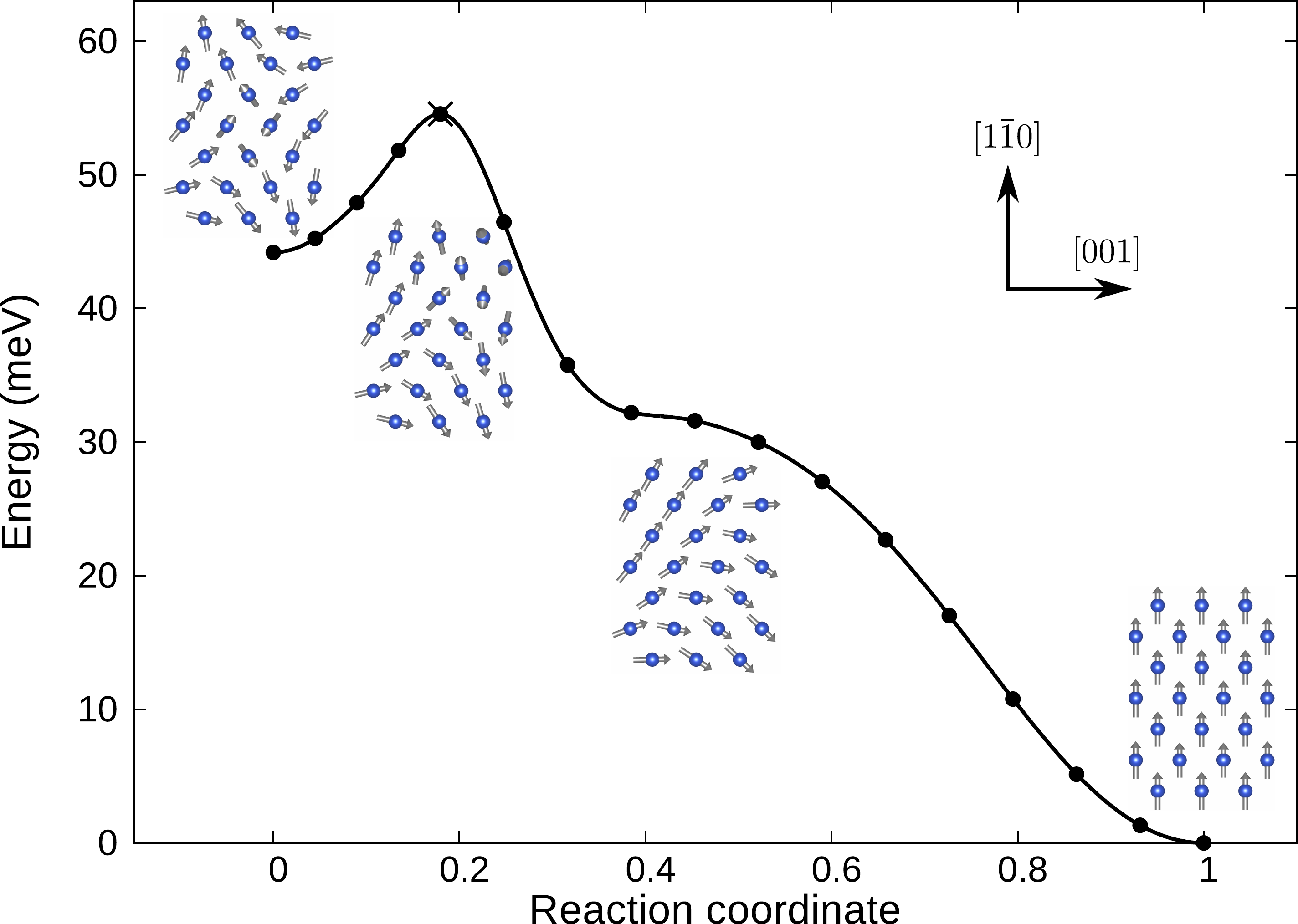}
\caption{
Energy along the minimum energy path between an antivortex metastable state and the collinear ground state 
of an Fe-island with $7\times 7$ atomic rows. The energy maximum along the path, which corresponds to a first order saddle point on the energy surface, is marked with a $\times$. The reaction coordinate is defined as the normalized displacement along the path. The orientation of crystallographic axes, $\left[1\bar{1}0\right]$ and $[001]$, are indicated with arrows. The anisotropy axis, $K_\parallel$, is oriented along the $\left[1\bar{1}0\right]$ direction. Insets show the direction and magnitude of the magnetic moments at the minimum energy configurations, at the saddle point and at another intermediate configuration.
}
\label{fig_ex}
\end{figure}

Ten minimization calculations were carried out starting from different, random orientations of the 24 magnetic moments.
In three of these calculations, a noncollinear state with an antivortex spin structure was found (see Fig.~\ref{fig_ex}).
The collinear ground state was found in the other seven 
calculations. 
The metastable antivortex state can be described as a symmetrical, saddlelike arrangement of the magnetic moments
in the center of the island. The total in-plane magnetization is 
zero, while the out-of-plane magnetization is nonzero mainly due to four magnetic moments near the center of the island which point out of the (110) plane. 
The identification of this magnetic state illustrates the power of the methodology described above, which is  
efficient because of the magnetic force theorem.

Magnetic antivortices have been identified before (see, for example,~\cite{waeyenberge_06}), but those are much larger than the one we found here. Typically, the formation of an antivortex state is closely related to the magnetostatic interaction, which is usually negligible on the nanoscale. Here, the antivortex state demonstrates the complex exchange interaction which is included in the NCAA model 
even though it contains only a few parameters.
The trimer example discussed above, see Fig.~\ref{fig:Trimer}, 
shows
that both parallel and antiparallel alignment of magnetic moments is possible in a system without anisotropy. 
Such a behavior cannot be obtained within a Heisenberg-type model unless additional phenomenological terms and
additional parameters are introduced in the Hamiltonian. 
In the NCAA model, complex noncollinear states appear quite naturally.

An NEB calculation was carried out to estimate the thermal stability of the 
antivortex 
state.
The 
MEP for the
transition from this state to the collinear ground state is shown in Fig.~\ref{fig_ex}.
The saddle-like excitation moves along the diagonal of the island towards one of the corners (the upper right corner in 
the insets of Fig.~\ref{fig_ex}) where it leaves the island. 
The activation energy for the transition from the metastable state to the ground state was calculated to be 10 meV, 
while for the reverse transition it is 
55 meV. 

\section{
Summary
}

We have derived in this paper a magnetic force theorem for the NCAA model. According to this theorem, derivatives of the total energy with respect to 
the
number of {\it d} electrons and magnitude of magnetic moments vanish 
when self-consistency has been reached.
As a result, the energy gradient with respect to the orientation of magnetic moments or, more generally, any adiabatic parameter, 
can be computed without having to repeat the self-consistency calculations. 
This theorem is of great practical importance as it significantly reduces the computational effort involved in finding  
magnetic states, 
calculating MEPs between the states, simulating 
spin dynamics and, more generally, 
navigating on the energy surface.

The theorem can also be used to calculate magnetic exchange parameters, $J_{ij}$, for an arbitrary, noncollinear ordering of magnetic moments, by using finite differences of the forces. In general, the exchange coupling depends on orientation of magnetic moments and 
cannot be reduced to a bilinear term only~\cite{bessarab_13c, costa_05}.
Thus the NCAA model can account for a complex, non-Heisenberg magnetic exchange coupling, which could play an important role in the formation of exotic, noncollinear states, such as the antivortex state found here. It may even be possible to form
magnetic skyrmions~\cite{heinze_11}
within the NCAA model.  


\begin{acknowledgments}
This work was supported by the Government of Russian Federation (Grant No. 074-U01), RFBR Grant No. 14-02-00102, the Icelandic Research Fund, 
and the Nordic-Russian Training Network for Magnetic Nanotechnology (NCM-RU10121). 
\end{acknowledgments}


\appendix

\section{Continued-fraction expansion for the off-diagonal elements of the Green function} \label{app_1}

In the original method due to Haydock~\cite{haydock_72,haydock_75}, the continued-fraction representation is obtained for the diagonal elements of the Green function. The recursion method, however, can 
also be applied to
the off-diagonal elements, $G_{\mu\nu}(\omega) = \left<x_\mu\right|G(\omega)\left|x_\nu\right>$, $\mu\ne\nu$, 
where $\left|x_\mu\right>$ and $\left|x_\nu\right>$ are the members of an initial basis set, 
and 
indices $\mu$, $\nu$ enumerate both atomic site and spin projection. 

In Haydock's method, a basis set is found that tridiagonalizes the Hamiltonian. All members of that basis set as well as matrix elements of the tridiagonal Hamiltonian are found systematically after specifying the first vector, $\left|y_1\right>$. The matrix element $\left<y_1\right|G(\omega)\left|y_1\right>$ is then expressed in terms of a continued fraction. The choice of $\left|y_1\right>$ in this procedure is arbitrary. 

In order to find $G_{\mu\nu}(\omega)$, a tridiagonalization is performed four times, where the starting vectors are:
\begin{align*}
\left|y_1^a\right>&=\frac{1}{\sqrt{2}}\left(\left|x_\mu\right>+\left|x_\nu\right>\right),\\
\left|y_1^b\right>&=\frac{1}{\sqrt{2}}\left(\left|x_\mu\right>-\left|x_\nu\right>\right),\\
\left|y_1^c\right>&=\frac{1}{\sqrt{2}}\left(\left|x_\mu\right>+\mathrm i\left|x_\nu\right>\right),\\
\left|y_1^d\right>&=\frac{1}{\sqrt{2}}\left(\left|x_\mu\right>-\mathrm i\left|x_\nu\right>\right).
\end{align*}
In each basis, matrix element $\tilde{G}_{11}^{\zeta}(\omega)=\left<y_1^{\zeta}\right|G(\omega)\left|y_1^{\zeta}\right>$, 
$\zeta = a,b,c,d$, is expanded in terms of continued fraction as described in Refs.~\cite{haydock_72,haydock_75}. 
Continued-fraction representation for the real and imaginary part of the nondiagonal element, $G_{\mu\nu}(\omega)$, 
is thus given by:
\begin{align}
\Re G_{\mu\nu}(\omega) &= \frac{1}{2}\left(\tilde{G}_{11}^a(\omega)-\tilde{G}_{11}^b(\omega)\right),\label{eq:re_g}\\
\Im G_{\mu\nu}(\omega) &= \frac{1}{2}\left(\tilde{G}_{11}^d(\omega)-\tilde{G}_{11}^c(\omega)\right).\label{eq:im_g}
\end{align}

\section{Partial-fraction expansion of the Green function} \label{app_2}

In the recursion method, matrix elements of the Green function are expressed in terms of continued fractions of the following kind:

\begin{equation}
C(\omega) = 
\cfrac{1}{\omega-a_1-\cfrac{\left|b_1\right|^2}{\omega-a_2-\cfrac{\cdots}{\cdots-\cfrac{\left|b_{2P-1}\right|^2}{\omega - a_{2P}}}}}
\label{eq:g_con_frac}
\end{equation}
The method developed in Ref.~\cite{uzdin_98} 
involves
systematically reducing the number of levels in the continued fraction, starting from the last level. Here, we briefly describe the idea of this method. Let us consider the last level of the continued fraction, Eq.~(\ref{eq:g_con_frac}):
\begin{equation}
\label{eq:f_1}
f^{(1)}(\omega) = \omega - a_{2P-1}-\frac{\left|b_{2P-1}\right|^2}{\omega - a_{2P}}.
\end{equation}

If $b_{2P-1}\ne0$, then the equation $f^{(1)}(\omega) = 0$ has two real roots, 
$q_1^{(2)}$ and $q_2^{(2)}$, $q_1^{(2)}<q_2^{(2)}$, which can be found either analytically or numerically. 
Then
$$
f^{(1)}(\omega) = \frac{\left(\omega-q_1^{(2)}\right)\left(\omega-q_2^{(2)}\right)}{\omega - q_1^{(1)}},
$$
where $q_1^{(1)}\equiv a_{2P}$.
As a result, the next level of the continued fraction acquires the form:
\begin{equation}
\label{eq:f_2}
f^{(2)}(\omega) = \omega - a_{2P-2}-\frac{\left|b_{2P-2}\right|^2\left(\omega-q_1^{(1)}\right)}{\left(\omega-q_1^{(2)}\right)\left(\omega-q_2^{(2)}\right)}=\omega - a_{2P-2}- \frac{p_1^{(2)}}{\omega-q_1^{(2)}}-\frac{p_2^{(2)}}{\omega-q_2^{(2)}},
\end{equation}
where $p_1^{(2)}$ and $p_2^{(2)}$ are defined as
$$
p_1^{(2)} = \left|b_{2P-2}\right|^2\frac{q_1^{(2)}-q_1^{(1)}}{q_1^{(2)}-q_2^{(2)}}, \qquad p_2^{(2)} = \left|b_{2P-2}\right|^2\frac{q_2^{(2)}-q_1^{(1)}}{q_2^{(2)}-q_1^{(2)}},
$$
Thus function $f^{(2)}(\omega)$ is represented in  
a
form which is analogous to Eq.~(\ref{eq:f_1}),
and the number of levels in the continued fraction is reduced by one. 
The same technique is used in order to successively eliminate all levels of $C(\omega)$. 
At each step $k$, finding zeros of $f^{(k)}(\omega)$ does not cause problems, because they are well separated.

\section{Proof of lemmas~(\ref{eq:lemma1}) and~(\ref{eq:lemma2})} \label{app_3}

In this appendix, we sketch the proof of the lemmas ~(\ref{eq:lemma1}) and~(\ref{eq:lemma2}), which can be written as
\begin{equation}
\label{eq:lemma1_app}
\frac{\partial\Tr{G(\omega)}}{\partial N_i} = -\frac{U_i}{2}\frac{\partial}{\partial \omega}\left(G_{ii}^{++}(\omega)+G_{ii}^{--}(\omega)\right),
\end{equation}
\begin{equation}
\label{eq:lemma2_app}
\begin{split}
\frac{\partial\Tr{G(\omega)}}{\partial M_i} &= \frac{U_i}{2}\frac{\partial}{\partial \omega}\left[\left(G_{ii}^{++}(\omega)-G_{ii}^{--}(\omega)\right)\cos{\theta_i}\right.\\
&+\left.\left(G_{ii}^{+-}(\omega)e^{i\phi_i}+G_{ii}^{-+}(\omega)e^{-i\phi_i}\right)\sin{\theta_i}\right].
\end{split}
\end{equation}

Due to the resolvent identity, we have:
\begin{align}
\frac{\partial G(\omega)}{\partial N_i} &= G(\omega)\cfrac{\partial H}{\partial N_i} G(\omega),\label{handy_1}\\
\frac{\partial G(\omega)}{\partial M_i} &= G(\omega)\cfrac{\partial H}{\partial M_i} G(\omega).\label{handy_2}
\end{align}
It follows directly from Eq.~(\ref{eq:AA_MF}) that 
\begin{align}
\left(\cfrac{\partial H}{\partial N_i}\right)_{kj}^{\alpha\beta} &= \frac{U_i}{2} \delta^{\alpha\beta}\delta_{ki}\delta_{ji},\label{dhn}\\ 
\left(\cfrac{\partial H}{\partial M_i}\right)_{kj}^{\alpha\beta} &= \frac{U_i}{2} \delta_{ki}\delta_{ji}\left[-\alpha\delta^{\alpha\beta}\cos\theta_i + \left(\delta^{\alpha\beta}-1\right)\sin\theta_i e^{-\alpha \mathrm i \phi_i}\right].\label{dhm}
\end{align}
Using Eqs.~(\ref{handy_1}), (\ref{dhn}) and invariance of a trace under cyclic permutations we get:
\begin{equation}
\label{eq:proof_1}
\begin{split}
\frac{\partial\Tr{G(\omega)}}{\partial N_i} &= \Tr G^2(\omega)\cfrac{\partial H}{\partial N_i}=-\Tr \frac{\partial G(\omega)}{\partial \omega}\cfrac{\partial H}{\partial N_i}=\\
&-\frac{U_i}{2}\left(\frac{\partial G^{++}_{ii}(\omega)}{\partial \omega}+\frac{\partial G^{--}_{ii}(\omega)}{\partial \omega}\right),
\end{split}
\end{equation}
where use was made of the following identity:
$$
\frac{\partial G(\omega)}{\partial \omega} = -G^2(\omega).
$$
Eq.~(\ref{eq:proof_1}) proves lemma~(\ref{eq:lemma1_app}). Lemma~(\ref{eq:lemma2_app}) is proven in the same way. We have:
\begin{equation}
\begin{split}
\frac{\partial\Tr{G(\omega)}}{\partial M_i} &= \Tr G^2(\omega)\cfrac{\partial H}{\partial M_i}=-\Tr \frac{\partial G(\omega)}{\partial \omega}\cfrac{\partial H}{\partial M_i}=\\
&
\frac{U_i}{2}\left[\left(\frac{\partial G^{++}_{ii}(\omega)}{\partial \omega}-\frac{\partial G^{--}_{ii}(\omega)}{\partial \omega}\right)\cos{\theta_i}\right.\\
&+\left.\left(\frac{\partial G_{ii}^{+-}(\omega)}{\partial \omega}e^{i\phi_i}+\frac{\partial G_{ii}^{-+}(\omega)}{\partial \omega}e^{-i\phi_i}\right)\sin{\theta_i}\right],
\end{split}
\end{equation}
which proves lemma~(\ref{eq:lemma2_app}).



\end{document}